\documentclass[trackchanges,preprint2]{aastex62}

\usepackage{savesym}
\savesymbol{tablenum}
\usepackage{siunitx}
\restoresymbol{SIX}{tablenum}
\usepackage{amsmath}
\usepackage{breqn}
\usepackage{verbatim}
\usepackage{multirow}
\usepackage{booktabs}

\DeclareSIUnit[space-before-unit=True]
\Msun{M_{\odot}}
\DeclareSIUnit[space-before-unit=True]
\Rsun{R_{\odot}}
\DeclareSIUnit[space-before-unit=True]
\Mearth{M_{\bigoplus}}
\DeclareSIUnit[space-before-unit=True]
\Gyr{Gyr}
\DeclareSIUnit[space-before-unit=True]
\Myr{Myr}
\DeclareSIUnit[space-before-unit=True]
\kyr{kyr}
\DeclareSIUnit[space-before-unit=True]
\yr{yr}
\DeclareSIUnit[space-before-unit=True]
\gal{gal}
\DeclareSIUnit[space-before-unit=True]
\AU{AU}

\newcommand{\vk}{v_{\rm kick}}
\newcommand{\vc}{v_{\rm circ}}

\newcommand{\Rroche}{R_{\rm{Roche}}}
\newcommand{\Mdisk}{M_{\mathrm{disk}}}

\graphicspath{{./}{Figures/}}

\submitjournal{ApJL}

\shorttitle{Planetesimal Tidal Disruption from a White Dwarf Kick}
\shortauthors{Akiba, McIntyre, \& Madigan}

\begin{document}

\title{Tidal Disruption of Planetesimals from an Eccentric Debris Disk Following a White Dwarf Natal Kick}

\author[0000-0002-0647-718X]{Tatsuya Akiba}
\affiliation{JILA and Department of Astrophysical and Planetary Sciences, CU Boulder, Boulder, CO 80309, USA}
\email{tatsuya.akiba@colorado.edu}

\author{Selah McIntyre}
\affiliation{Department of Chemistry, University of Colorado, Boulder, CO 80309, USA}

\author[0000-0002-1119-5769]{Ann-Marie Madigan}
\affiliation{JILA and Department of Astrophysical and Planetary Sciences, CU Boulder, Boulder, CO 80309, USA}

\begin{abstract}

The surfaces of many white dwarfs are polluted by metals, implying a recent accretion event. The tidal disruption of planetesimals is a viable source of white dwarf pollution and offers a unique window into the composition of exoplanet systems. The question of how planetary material enters the tidal disruption radius of the white dwarf is currently unresolved. Using a series of $N$-body simulations, we explore the response of the surrounding planetesimal debris disk as the white dwarf receives a natal kick caused by anisotropic mass loss on the asymptotic giant branch. We find that the kick can form an apse-aligned, eccentric debris disk in the range $\SIrange{30}{240}{\AU}$ which corresponds to the orbits of Neptune, the Kuiper Belt, and the scattered disk in our solar system. In addition, many planetesimals beyond $\SI{240}{\AU}$ flip to counter-rotating orbits. Assuming an isotropic distribution of kicks, we predict that approximately $80\%$ of white dwarf debris disks should exhibit significant apsidal alignment and fraction of counter-rotating orbits. The eccentric disk is able to efficiently and continuously torque planetesimals onto radial, star-grazing orbits. We show that the kick causes both an initial burst in tidal disruption events as well as an extended period of $\SI{100}{\Myr}$ where tidal disruption rates are consistent with observed mass accretion rates on polluted white dwarfs.

\end{abstract}


\section{Introduction} \label{sec:intro}

White dwarfs are the remnants of stars with main-sequence masses below $\SI{8}{\Msun}$, which constitute an estimated $97\%$ of stars in our Galaxy --- including our Sun \citep{Fontaine2000}. White dwarfs are extremely dense bodies with masses comparable to the Sun, despite their sizes being closer to that of the Earth \citep{Schatzman1958}. Most white dwarfs should have a core of carbon and oxygen surrounded by a thin, outer layer of hydrogen and helium \citep{Burbidge1954}. However, an estimated $25$ to $50\%$ of observed white dwarfs show signs of metals such as calcium, magnesium, iron, and silicon in their spectra \citep[e.g.,][]{Zuckerman2003, Zuckerman2010, Koester2014}. Metal polluted white dwarfs are found with a wide range of effective temperatures $\SIrange{3000}{25000}{\K}$ \citep[see][]{Farihi2016} with estimated cooling ages as old as $\SI{10}{\Gyr}$ \citep{Elms2022}.

The presence of metals in white dwarf spectra is commonly attributed to the active accretion of planetary material following a tidal disruption event \citep[e.g.,][]{Debes2002, Jura2003, Wang2019, Brouwers2022}, given that any surface metals should sink to the core on a timescale much shorter than the white dwarf age through gravitational settling \citep[see][]{JuraYoung2014}. The radius at which this tidal disruption occurs is the white dwarf's Roche radius $\approx \SI{1}{\Rsun}$ \citep[e.g,][]{Li1998, Davidsson1999, BearSoker2013, Veras2014,  Barber2016, Veras2021}. Once a planetesimal is tidally disrupted, the bound planetary debris forms a circumstellar disk which produces an observable infrared excess \citep[e.g.,][]{Jura2003, Jura2007}, and many white dwarfs have been found with surrounding planetary material in this manner \citep{Farihi2016}. The composition of tidally disrupted planetesimals can be inferred from metal abundances on polluted white dwarfs which provides a unique avenue for the study of exoplanet compositions \citep[e.g.,][]{Zuckerman2007, Zuckerman2010, Koester2014}.

Estimated mass accretion rates are high and challenging to explain \citep[e.g.,][]{Brouwers2022}. Time-averaged accretion rates inferred from helium white dwarfs are typically $\SI{e9}{\g\per\s}$, whereas instantaneous accretion rates measured from hydrogen white dwarfs are around $\SI{e7}{\g\per\s}$ \citep{Farihi2012, Hollands2018, Blouin2022}. The highest rates inferred from helium and hydrogen white dwarfs are approximately $\SI{e11}{\g\per\s}$ and $\SI{e9}{\g\per\s}$, respectively \citep[e.g.,][]{Dufour2012, Gansicke2012, Xu2013, Farihi2016b, Cunningham2022}. There are many proposed mechanisms for the tidal disruption of planetesimals including perturbations due to secular resonances \citep{Smallwood2018}, binary stellar companions \citep[e.g.,][]{BonsorVeras2015, Hamers2016}, secular instabilities triggered by planetary engulfment \citep{Petrovich2017}, and mass loss on the asymptotic giant branch \citep{Reimers1975, Bloecker1995, Bonsor2011}. Exomoons \citep[e.g.,][]{Trierweiler2022} and planets \citep[e.g.,][]{Frewen2014} have also been proposed as potential sources of pollution in addition to asteroids and comets.

As a main-sequence star with mass below $\SI{8}{\Msun}$ runs out of hydrogen in its core, it turns into a red giant star to undergo subsequent fusion of heavier elements \citep{Iben1967}. On the asymptotic giant branch, the outer layers of the star become unbound and about half of the stellar mass is lost before the core collapses into a white dwarf \citep{Auer1965, Fusi-Pecci1976}. When this mass ejection occurs anisotropically, a natal kick is imparted on the white dwarf upon its formation \citep{Fellhauer2003}. The kick magnitude is expected to be $\SIrange{1}{3}{\km\per\s}$ \citep[e.g.,][]{Fregeau2009, El-Badry2018,Hamers2019}; the direction of the kick with respect to the planetesimal disk plane is unknown. \citet{Stone2015} considered the effect of this natal kick on exo-Oort clouds. The kick maps many comets onto radial, plunging orbits which produces a temporary burst of tidal disruption events. However, their Monte Carlo approach followed post-kick orbits on a short timescale without self-gravity, and hence does not apply to the cooler population of metal-polluted white dwarfs. In this letter, we show that the white dwarf natal kick results in the formation of an apse-aligned, eccentric debris disk of planetesimals which produces tidal disruption events at a rate consistent with observed mass accretion rates for $\SI{100}{\Myr}$.

\section{Eccentric Debris Disk Formation and Stability} \label{sec:analytics}

When a gravitational recoil kick is imparted on a supermassive black hole, the surrounding stellar orbits in a nuclear star cluster can form an eccentric, apse-aligned disk \citep{Akiba2021, Akiba2023}, and these eccentric disks exhibit stellar tidal disruption rates as high as $\SI{0.1}{\per\yr\per\gal}$, three to four orders of magnitude higher than rates expected from isotropic distributions \citep{Madigan2018}. The dynamics work on all scales in a near-Keplerian system and thus are directly applicable to planetesimals surrounding white dwarfs following the impartment of a natal kick. To quantify apsidal alignment, we make use of the eccentricity vector

\begin{equation}
\vec{e} = \frac{\vec{v} \times \vec{j}}{GM_{*}} - \hat{r} \ ,
\end{equation}

\noindent where $\vec{v}$ is the velocity vector, $\vec{j}$ is the (specific) angular momentum vector, $M_{*}$ is the white dwarf mass, and $\hat{r}$ is the unit position vector. The eccentricity vector points from the apoapsis to the periapsis of a given orbit, with a magnitude equal to the scalar eccentricity.

The \textit{mean eccentricity vector} is a measure of apsidal alignment defined by

\begin{equation}
\langle \vec{e} \rangle \equiv \frac{\sum_i^N \vec{e}_i}{N} \ , 
\label{eqn:mean_ecc}
\end{equation}

\noindent where $\vec{e}_i$ is the eccentricity vector of the $i$-th planetesimal and $N$ is the number of planetesimals considered. When the white dwarf experiences an in-plane kick, planetesimals on initially circular orbits will align their eccentricity vectors such that

\begin{equation}
    |\langle \vec{e} \rangle| = \frac{3}{2} \frac{\vk}{\vc} \ ,
\label{eqn:mean_ecc_vec}
\end{equation}

\noindent where $\vk$ is the natal kick speed and $\vc$ is the initial circular speed of the planetesimals. Apsidal alignment is strongest when $| \langle \vec{e} \rangle | = 1$. From Equation \ref{eqn:mean_ecc_vec}, this occurs at a \textit{characteristic radius}
\begin{equation}
    r_c = \frac{4GM_{*}}{9 \vk^2} \ .
\label{eqn:aps_align_radius}
\end{equation}

\noindent For $M_{*} = \SI{0.6}{\Msun}$ and $\vk = \SI{1}{\km\per\s}$, this radius occurs at $r_c = \SI{240}{\AU}$. For $\vk = \SI{3}{\km\per\s}$, $r_c = \SI{30}{\AU}$.

In the solar system, an orbital distance of $\SI{30}{\AU}$ corresponds to that of Neptune and the Kuiper Belt, a dynamically rich region of space that includes both kinematically cold and hot primordial planetesimal populations intermixed with those in orbital resonance with Neptune \citep[e.g.,][]{Jewitt1993, Malhotra2019}. A distance of $\SI{240}{\AU}$ corresponds to that of the scattered disk, a population of planetesimals on eccentric orbits with periapses that bring them into contact with Neptune's orbit \citep{Duncan1997Sci,Vokrouhlicky2019}. We note that this comparison does not take into account orbital expansion due to the star's mass loss \citep[e.g.,][]{Veras2013}. The existence of Kuiper Belt or scattered disk-like structures in exoplanet systems has been inferred from observations \citep[e.g.,][]{Geiler2019, Wyatt2020}; disks of icy bodies in the outskirts of planetary systems should be common.

Dynamical stability in eccentric disks comes about via mutual gravitational torques between orbiters \citep{Madigan2018}. When a planetesimal precesses ahead of the eccentric disk, its orbit is negatively torqued by the disk, and its angular momentum decreases which in turn increases its scalar eccentricity. This change in eccentricity works to slow down the orbit's precession and allows the rest of the disk to catch up to it. The opposite is true for an orbit that lags behind the disk: it feels a positive torque which circularizes the orbit and increases its precession speed. In this way, the eccentric disk maintains its apsidal alignment as individual orbits undergo oscillations in precession speeds and eccentricity. It is the latter oscillation in eccentricity that causes an enhancement in the rate of tidal disruption events as strong mutual torques throw planetesimals onto radial, star-grazing orbits.

\section{Numerical Simulations}
\label{sec:sims}

\subsection{Initial Setup} \label{ss:setup}

We use the open-source, $N$-body simulation package \texttt{REBOUND} \citep{Rein2012}, with code units of $G = 1$, $M_{*} = 1$, and  $r_c = 1$ where $r_c$ is the \textit{characteristic radius} given by Equation \ref{eqn:aps_align_radius}. For concreteness, we translate to physical units assuming a white dwarf mass of $M_{*} = \SI{0.6}{\Msun}$ and a kick velocity of $\vk = \SI{1}{\km\per\s}$ throughout the letter. This sets $r_c = \SI{240}{\AU}$. However, the dynamics are completely scalable; the presented results can be applied to a different $M_*$ or $\vk$ by scaling lengths by $r_c \propto M_*/\vk^2$ (see Equation \ref{eqn:aps_align_radius}) and timescales by $t \propto r_c^{3/2}$.

\begin{figure}[t!]
\centering
\includegraphics[width=\linewidth]{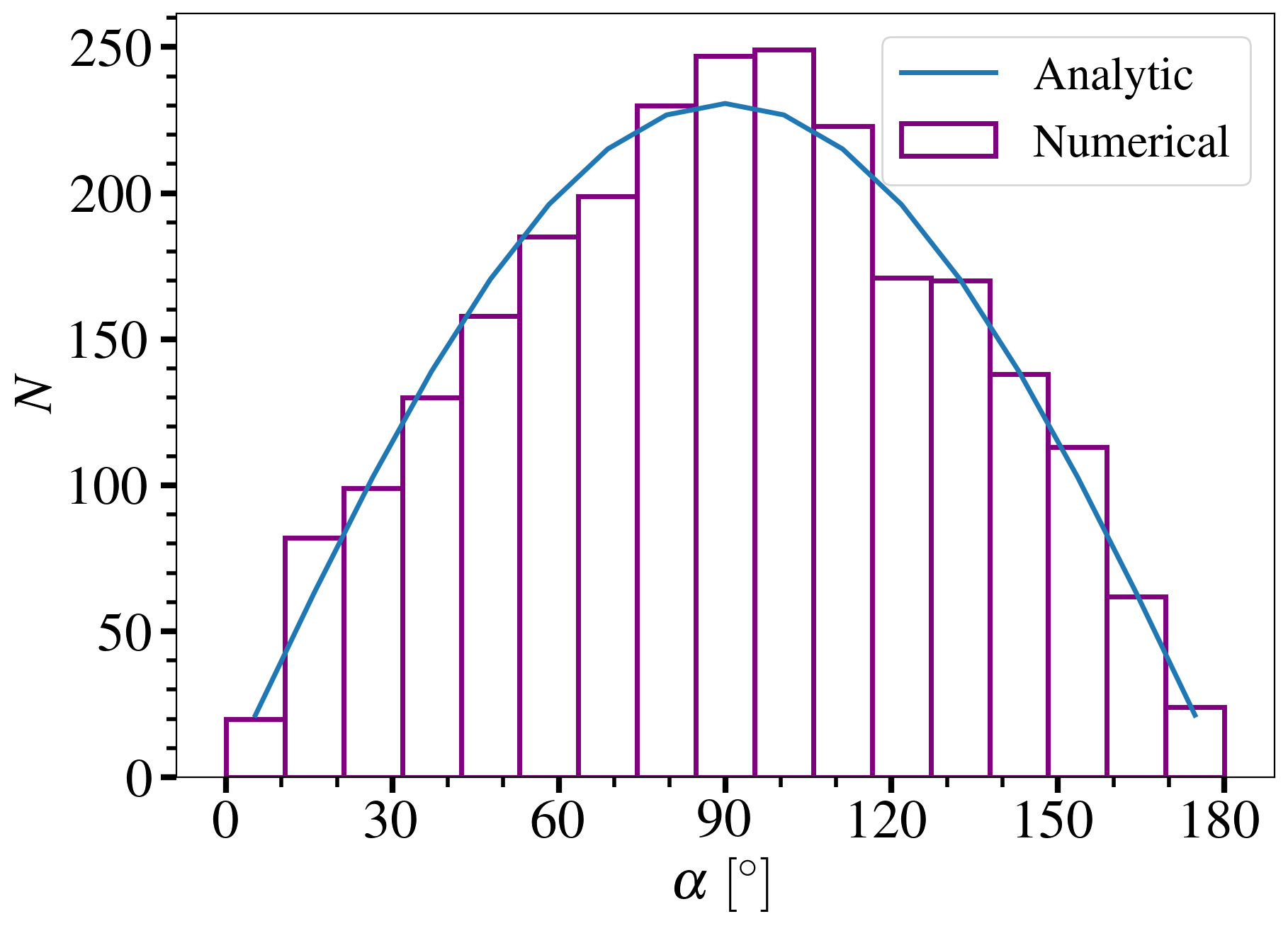}
\caption{A histogram of kick angles with respect to the orbital angular momentum axis. $\alpha = \SI{0}{\degree}$ or $\SI{180}{\degree}$ indicates an out-of-plane kick where the kick is parallel or anti-parallel to the angular momentum vector, whereas $\alpha = \SI{90}{\degree}$ means that the kick is in the planetesimal disk plane. The analytic expectation assuming an isotropic distribution is shown with the blue, solid line and the numerical results of sampling 2500 times are shown in purple bars. The most likely natal kick direction is in the planetesimal disk plane.}
\label{fig:alpha_hist}
\end{figure}

To study the instantaneous post-kick distribution, we initialize $N = \num{5e4}$ massless planetesimals in an axi-symmetric, thin disk spanning four orders of magnitude in semi-major axis space. We use a surface density profile $\Sigma \propto a^{-1.5}$ as motivated by \citet{Hayashi1981}. The number of planetesimals was chosen to have a sufficient number density out to Oort cloud distances of $a \approx \SI{e3}{\AU}$. Inclination is Rayleigh-distributed with scale parameter $\sigma = \SI{3}{\degree}$ while longitude of periapsis and mean anomaly are uniformly distributed in $[0, 2 \pi)$. We perform 2500 simulations in which the white dwarf instantaneously gains a velocity of $\vk = \SI{1}{\km\per\s}$ with respect to its initial frame of reference at time $t=0$, allowing orbits to impulsively respond. We define the planetesimal disk plane to be in the $x$-$y$ plane with the angular momentum vector pointing in the $+z$-direction. We further define the $+x$-direction to be the direction of the in-plane component of the kick. We randomly sample the kick angle with respect to the disk plane from an isotropic distribution. The resulting distribution of kick angles is shown in Figure \ref{fig:alpha_hist}. We note that $\alpha$ is measured with respect to the $+z$-direction. Since $\alpha$ is uniform in cosine, in-plane kicks are more likely than out-of-plane kicks.

\begin{figure*}[t!]
\centering
\includegraphics[width=\linewidth]{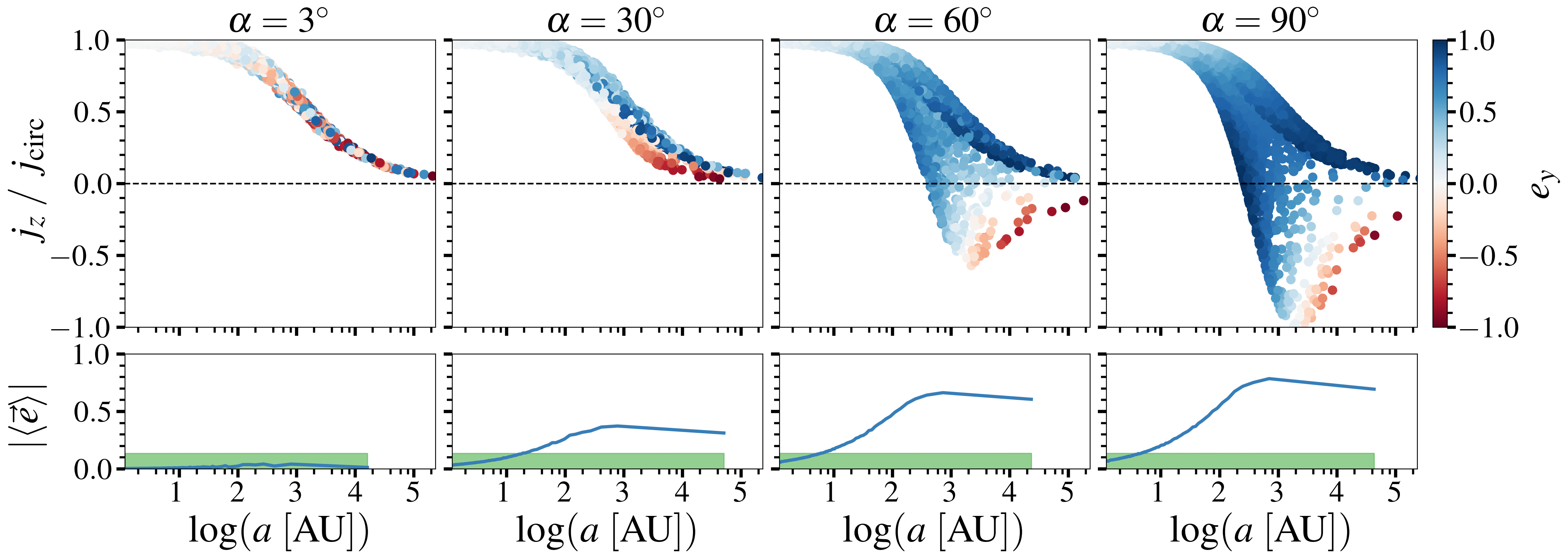}
\caption{A comparison of the planetesimals' orbital angular momenta and apsidal alignment as a function of semi-major axis at $t=0$ post-kick ranging from $\alpha = \SIrange[range-units=repeat]{3}{90}{\degree}$. (\textit{Top:}) The $z$-component of angular momentum normalized by the circular angular momentum, $j_{\rm{circ}} = \sqrt{GM_*a}$. Orbits above the dashed line are prograde and those below are retrograde. Orbits are more eccentric closer to the dashed line and are circular when $|j_z/j_{\rm{circ}}| = 1$. The planetesimal orbits are color-coded by the $y$-component of the eccentricity vector. (\textit{Bottom:}) The magnitude of the mean eccentricity vector, $\langle \vec{e} \rangle$. The green, shaded region indicates the noise floor above which alignment is statistically significant at the $3\sigma$-level. Retrograde orbits and apsidal alignment both emerge as the kick tends toward an in-plane direction.}
\label{fig:sma_ang_mom_mean_ecc}
\end{figure*}

\subsection{Structure of Eccentric Debris Disks}

In Figure \ref{fig:sma_ang_mom_mean_ecc}, we compare the distributions of orbital angular momenta and apsidal alignment of planetesimals as a function of semi-major axis at $t=0$ post-kick for simulations with different kick angles with respect to the planetesimal disk. We opt to show kick angles in the range $\alpha = \SIrange[range-units=repeat]{0}{90}{\degree}$ since these results are completely symmetric about $\alpha = \SI{90}{\degree}$; the distributions for $\alpha = \SI{60}{\degree}$ and $\SI{120}{\degree}$ look identical, for instance. For each simulation, we show the distribution of the $z$-component of angular momentum in the top panel. The plot is color-coded by the $y$-component of the eccentricity vector which is the direction apsidal alignment is expected with a kick in the $+x$-direction. In the bottom panel for each simulation run, we show the distribution of the magnitude of the mean eccentricity vector, $| \langle \vec{e} \rangle |$ as defined in Equation \ref{eqn:mean_ecc}. $| \langle \vec{e} \rangle | = 0$ when the disk is axi-symmetric and deviates toward unity as the disk becomes apse-aligned.

\begin{figure*}[t!]
\centering
\includegraphics[width=\linewidth]{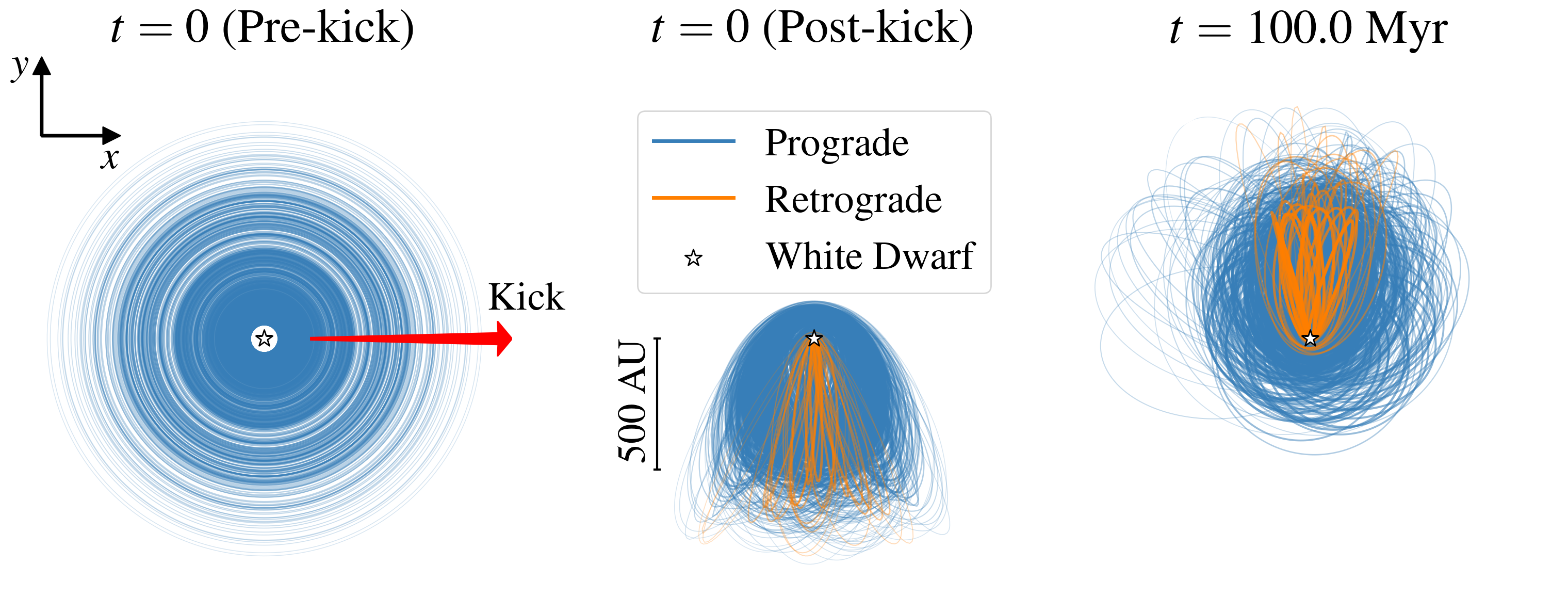}
\caption{Snapshots of planetesimal orbits for the $\Mdisk = \SI{20}{\Mearth}$ simulation at (\textit{left:}) $t=0$ pre-kick, (\textit{center:}) $t=0$ post-kick, and (\textit{right:}) $t = \SI{100}{\Myr}$. Initially, every planetesimal has a circular, prograde orbit. The kick forms an eccentric debris disk which coherently precesses while maintaining apsidal alignment.}
\label{fig:orbit_plot}
\end{figure*}

For kicks that are nearly out-of-plane, planetesimals remain on prograde orbits and their eccentricities increase with semi-major axis. There are no patterns in the distribution of $e_y$ and $| \langle \vec{e} \rangle | = 0$ throughout the disk. When $\alpha \geq \SI{30}{\degree}$, two dynamically important structures are observed: 1. strong apsidal alignment in the $+y$-direction, and 2. a significant \textit{retrograde} population of planetesimals beyond $\SI{240}{\AU}$. The retrograde planetesimal population emerges at large semi-major axes as the initial speed of the planetesimals becomes smaller than the natal kick speed. In the reference frame of the kicked white dwarf, the planetesimal velocity can flip such that the post-kick orbit is retrograde with respect to the initial angular momentum axis of the disk. In particular, the in-plane kick case shows a prograde population of planetesimals that is entirely apse-aligned in the $+y$-direction, a retrograde population at $a > \SI{240}{\AU}$ that is aligned in the same direction, and a retrograde population at $a > \SI{2000}{\AU}$ that is aligned in the opposite direction. Apsidal alignment is statistically significant throughout the debris disk, and the retrograde fraction is significant beyond $\SI{240}{\AU}$. It should be emphasized that at least one retrograde planetesimal is found in approximately $\SI{80}{\%}$ of our simulations. Apsidal alignment and retrograde planetesimals in white dwarf systems are thus natural consequences of natal kicks.

\subsection{Eccentric Debris Disk Evolution and the Tidal Disruption of Planetesimals}

To investigate the burst of tidal disruption events following the kick, we redistribute $N = \num{5e4}$ massless planetesimals in the region $a = \SIrange{240}{2400}{\AU}$ where apsidal alignment is the strongest. This is done in order to avoid issues with small number statistics in the detection of tidal disruption events. For the most probable case of an in-plane kick, we randomly select a subset of $N=\num{400}$ planetesimals in the range $a = \SIrange{240}{480}{\AU}$ and switch them from massless test particles to massive, self-interacting particles to study the eccentric debris disk's long-term evolution. We use \texttt{REBOUND}'s \texttt{IAS15} integrator, a high-order, non-symplectic integrator with adaptive time-stepping \citep{ReinSpiegel2015}. The low $N$ and narrow range of semi-major axes explored are in part due to the computational limitations of \texttt{IAS15}, but high-accuracy integration is critical in studying eccentric disk evolution\footnote{We also tested \texttt{WHFast}, a low-order, symplectic integrator \citep{Rein2015}. Our results showed that \texttt{WHFast} runs significantly deviate from \texttt{IAS15} runs of the same disk mass with the former exhibiting total energy and angular momentum errors of order unity within a $\unit{\Myr}$. The energy and angular momentum errors are $< \num{e-10}$ even for long-term integrations up to $\SI{100}{\Myr}$ using \texttt{IAS15}. Due to the high-resolution required to accurately integrate eccentric orbits near their periapses accompanied by strong mutual torques between orbits, high-order integrators such as \texttt{IAS15} are required to fully study the evolution of an eccentric disk.}. 

We vary the total disk mass in the range $\Mdisk = \SIrange{1}{100}{\Mearth}$ to study the effect of disk mass on the tidal disruption rate. Tidal disruption events are detected by comparing the periapsis distance, $r_- = a (1-e)$ with the Roche radius defined to be $\Rroche \equiv \SI{1}{\Rsun}$ at each time step. When $r_- < \Rroche$, the planetesimal is considered tidally disrupted. Each massive simulation is stopped at $t = \SI{10}{\Myr}$. The only exception is the simulation with $\Mdisk = \SI{20}{\Mearth}$ which is run until $t = \SI{100}{Myr}$ to show the long-term enhancement of tidal disruption rates. The simulation run is stopped at $\SI{100}{\Myr}$ due to two computational limitations of our low $N$ setup: 1. the number of tidally disrupted planetesimals becomes comparable to $N$, and 2. the two-body relaxation timescale \citep{Rauch1996}, which is proportional to $N$ for a given disk mass, becomes comparable to the simulation time \citep[see][]{Madigan2018b}.

\begin{figure*}[t!]
\centering
\includegraphics[width=\linewidth]{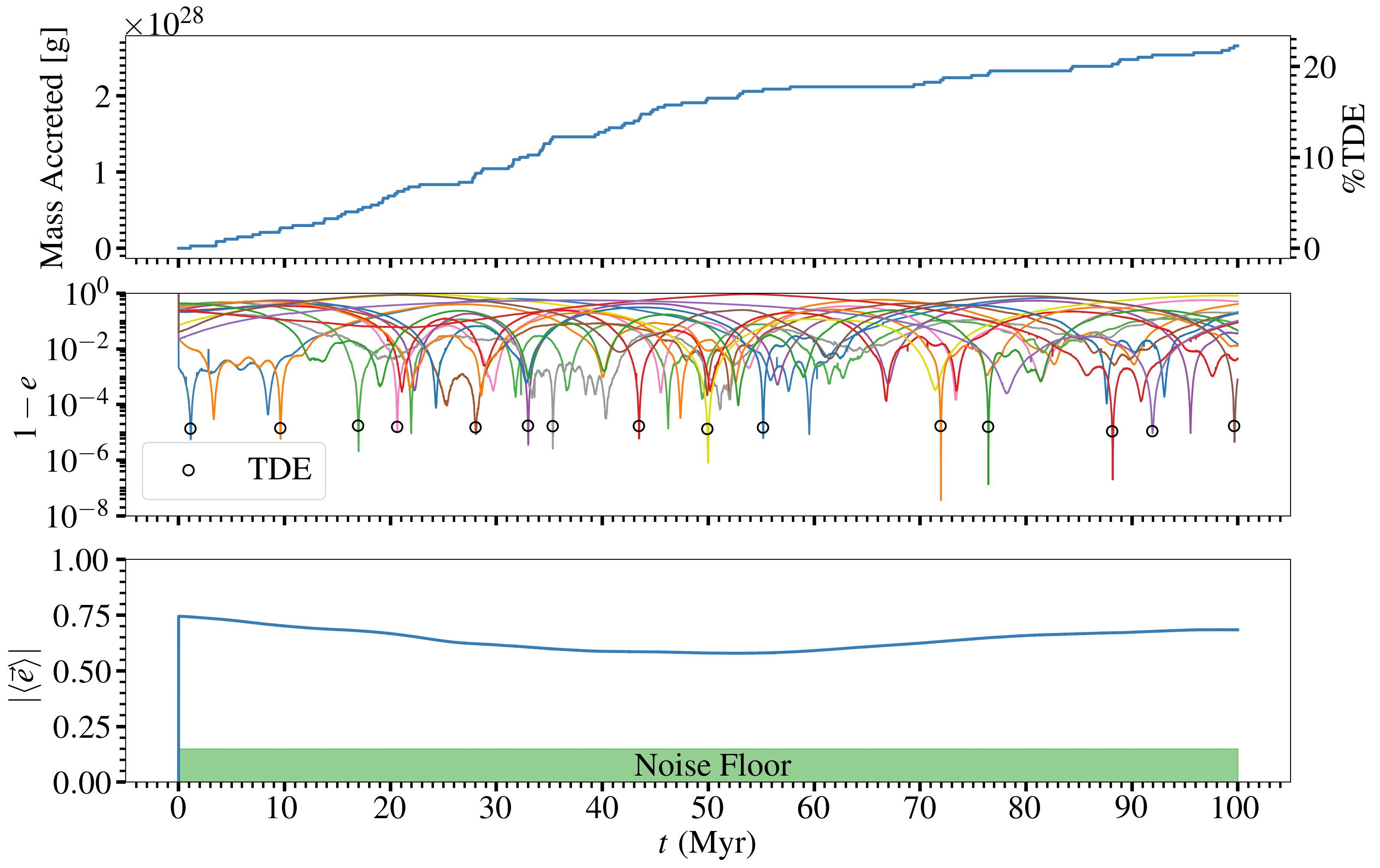}
\caption{Evolution of tidal disruption rates and apsidal alignment for the $\Mdisk = \SI{20}{\Mearth}$ simulation. (\textit{Top:}) The cumulative mass accreted by the white dwarf from tidal disruption events over time. The axis on the right shows the fraction of the debris disk that is tidally disrupted. (\textit{Center:}) Eccentricity evolution of a small sample of planetesimals that become tidally disrupted. The quantity, $1-e$, is plotted on the $y$-axis using a logarithmic scale. The time at which tidal disruption is detected is denoted by an open black circle. (\textit{Bottom:}) The evolution of the magnitude of the mean eccentricity vector. The green shaded region indicates the noise floor above which apsidal alignment is statistically significant at the $3\sigma$-level.}
\label{fig:tde_evol}
\end{figure*}

In Figure \ref{fig:orbit_plot}, we show planetesimal orbit projections at $t=0$ pre-kick, $t=0$ post-kick, and $t = \SI{100}{\Myr}$ after the kick for the $\Mdisk = \SI{20}{\Mearth}$ simulation. Initially, the pre-kick orbits are all circular and prograde. Post-kick, the orbits' eccentricity vectors coherently point in the $+y$-direction. We also see the emergence of retrograde orbits which are apsidally aligned in the same direction. For $\SI{100}{\Myr}$, the alignment of eccentricity vectors is maintained while the eccentric disk coherently precesses in the prograde direction. We perform a quick estimation of the fraction of kicked white dwarfs that tidally disrupt a planetesimal. At $t=0$ post-kick, we compare the new periapsis distances of the redistributed $\num{5e4}$ massless planetesimals in the range $a = \SIrange{240}{2400}{\AU}$ to $\Rroche$. Of the massless simulations with isotropically-distributed kicks, we find that 10\% of white dwarfs immediately tidally disrupt at least one planetesimal upon receiving a kick of $\SI{1}{\km\per\s}$. This fraction increases to 16\% if a $\SI{3}{\km\per\s}$ kick is assumed instead. Interestingly, \citet{Stone2015} predict a similar fraction of comets that tidally disrupt in exo-Oort clouds. However, this fraction is notably lower than the expected $25$ to $50$\% of older white dwarfs that are metal-polluted. We show in the following section that eccentric debris disks can increase this fraction by inducing tidal disruption events at later times through strong mutual torques between planetesimals.

\subsection{Long-term Enhancement of Tidal Disruption Rates in Eccentric Disks}

In the top panel of Figure \ref{fig:tde_evol}, we show the cumulative mass accreted by the white dwarf from planetesimal tidal disruption events during the $\Mdisk = \SI{20}{\Mearth}$ simulation. We note that we do not include a treatment of the accretion process; we are inherently assuming that a planetesimal, once tidally disrupted, is entirely accreted by the white dwarf on orbital timescales $t_{\rm{orb}} \approx \unit{\kyr}$. We see that the eccentric debris disk is able to produce tidal disruption events for an extended period of $\SI{100}{\Myr}$ following the impartment of the kick. This particular disk mass simulation implies a very high accretion rate of $\SI{e12}{\g\per\s}$. Except for a few periods or dormancy (e.g., $t = \SIrange{58}{69}{\Myr}$), at least one tidal disruption event is observed every few $\unit{\Myr}$ and hence the expected time-averaged accretion rate is at least $\SI{e9}{\g\per\s}$ throughout the simulation. In addition, due to the significant retrograde fraction induced by the kick, we find that approximately 40\% of tidally disrupted planetesimals are on retrograde orbits when they become tidally disrupted. If the initial disk angular momentum axis is the same as that of the star's rotation axis, this mechanism would predict a significant fraction of white dwarf systems to have a circumstellar debris disk following a tidal disruption event that is retrograde with respect to the white dwarf spin axis.

In the center panel of Figure \ref{fig:tde_evol}, we plot the eccentricity evolution of a small sample of planetesimals that become tidally disrupted. While the initial jump to high eccentricities at $t=0$ is due to the natal kick, the kick alone is unable to tidally disrupt these planetesimals. The subsequent evolution to higher eccentricities is caused by the strong mutual torques between planetesimals within the eccentric disk (see Section \ref{sec:analytics}). The bottom panel of Figure \ref{fig:tde_evol} shows the apsidal alignment evolution. Statistically-significant alignment is maintained throughout the simulation. We note that the oscillation in $| \langle \vec{e} \rangle |$ is caused by the periodic alignment and misalignment between the eccentric disk and planetesimals near the outer edge of the disk that break off and precess in the retrograde direction \citep[see][]{Madigan2018, Akiba2021}. This figure shows that the oscillation is correlated with bursts of tidal disruption events; as the retrograde-precessing planetesimals come back into alignment with the eccentric disk, mutual gravitational torques increase the amplitude of the eccentricity oscillations which consequently increase tidal disruption rates. The white dwarf natal kick causes an initial burst of tidal disruption events followed by an extended $\SI{100}{\Myr}$ period where tidal disruption rates from the eccentric debris disk are consistent with some of the highest mass accretion rates observed. We note that $t = \SI{100}{\Myr}$ corresponds to polluted white dwarfs on the warmer end with temperatures around $\SI{20000}{\K}$ \citep[see][]{Farihi2016}; a higher $N$ simulation is needed to extend this study to white dwarfs with cooling ages $\unit{Gyr}$ or older.

\subsection{Tidal Disruption Rate Dependence on Eccentric Debris Disk Mass}

\begin{figure}[t!]
\centering
\includegraphics[width=\linewidth]{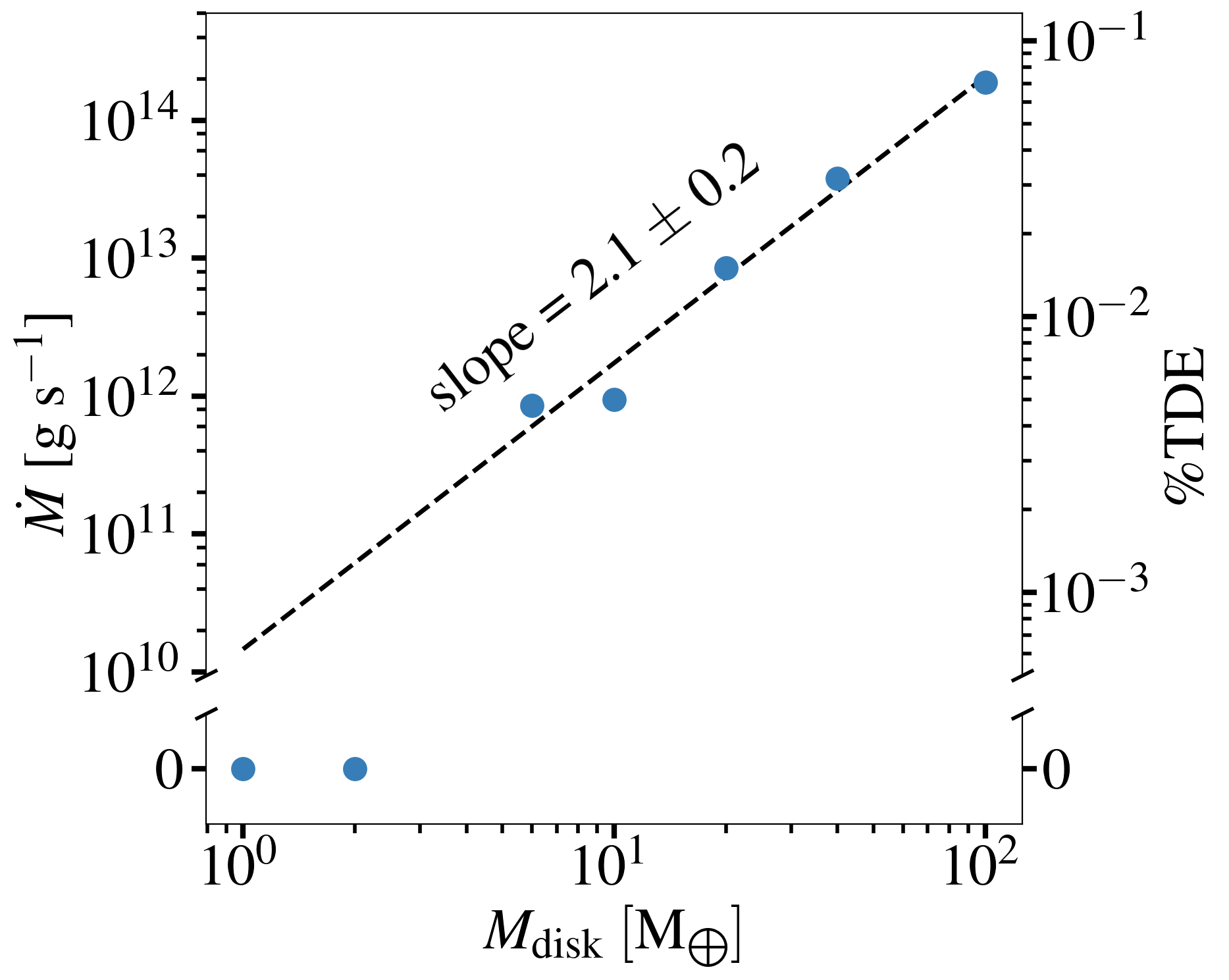}
\caption{The average mass accretion rate, $\dot{M}$ from tidal disruption events as a function of the debris disk mass, $\Mdisk$. The average is calculated from the first $\SI{10}{\Myr}$ of each simulation. The axis on the right shows the corresponding fraction of the debris disk that is tidally disrupted in each case. For simulations with at least one tidal disruption event detection, the results are consistent with a power-law dependence $\dot{M} \propto (\Mdisk)^2$.}
\label{fig:tde_comp}
\end{figure}

The expected planetesimal tidal disruption rate has a steep dependence on the mass of the eccentric debris disk. The following analysis assumes that the accretion process happens on the orbital timescale $t_{\rm{orb}} \approx \unit{\kyr}$ which is much shorter than the \textit{secular timescale} defined by $t_{\rm{sec}} \equiv (M_*/\Mdisk)~t_{\rm{orb}}$. The eccentricity oscillations that cause tidal disruption events happen on the secular timescale \citep{Madigan2018}, so the tidal disruption rate $\dot{N}_{\rm{TDE}} \propto t_{\rm{sec}}^{-1} \propto \Mdisk$. The mass accretion rate is estimated as $\dot{M} = m \dot{N}_{\rm{TDE}}$ where $m$ is the planetesimal mass. In our simulations, we fix $N$ and vary $\Mdisk$, so $m \propto \Mdisk$. Hence, we analytically predict $\dot{M} \propto (\Mdisk)^2$.

In Figure \ref{fig:tde_comp}, we show the estimated mass accretion rate from tidal disruption events for each of our massive disk simulations. For simulations with at least one tidal disruption event detected, the correlation is well-modeled by our analytic prescription $\dot{M} \propto (\Mdisk)^2$. The $\Mdisk = \num{1}$ and $\SI{2}{\Mearth}$ simulations detect no tidal disruption events in the first $\SI{10}{\Myr}$, but this is likely due to the low $N$ and short simulation time. For instance, $\Mdisk = \SI{1}{\Mearth}$ implies a secular timescale of $t_{\rm{sec}} \approx \SI{100}{\Myr}$. If instead, we extrapolate the $\dot{M} \propto (\Mdisk)^2$ scaling to lower debris disk masses, we predict that typical instantaneous accretion rates of $\SI{e7}{\g\per\s}$ on hydrogen white dwarfs are reproduced for any eccentric debris disk with mass above $\SI{0.03}{\Mearth}$. The highest measured instantaneous accretion rates of $\SI{e9}{\g\per\s}$ are predicted for $\Mdisk > \SI{0.3}{\Mearth}$. Finally, the highest time-averaged accretion rates inferred from helium white dwarfs of $\SI{e11}{\g\per\s}$ require debris disk masses of at least a few $\unit{\Mearth}$.

\section{Discussion} \label{sec:discussion}

We perform numerical simulations to explore the formation of an eccentric debris disk following a white dwarf natal kick. We run 2500 massless simulations with isotropically-distributed kick angles to estimate the fraction of white dwarf debris disks with significant apsidal alignment, retrograde fraction, and planetesimal tidal disruption rates. For the most likely case of an in-plane kick, we follow the evolution of the eccentric disk and tidal disruption rates by performing a series of massive $N$-body simulations. Our main findings are:

\begin{enumerate}
    \item An eccentric debris disk and a retrograde planetesimal population are expected for approximately $80 \%$ of planetesimal debris disks. The retrograde population forms beyond $\SI{240}{\AU}$ for a white dwarf natal kick of $\SI{1}{\km\per\s}$.
    \item For a $\SI{1}{\km\per\s}$ kick, $10\%$ of white dwarfs are expected to immediately tidally disrupt a planetesimal. This fraction increases to $16\%$ for a $\SI{3}{\km\per\s}$ kick. This is consistent with earlier results of \citet{Stone2015}.
    \item The eccentric debris disk hosts planetesimal tidal disruption rates consistent with observed mass accretion rates for an extended $\SI{100}{\Myr}$ period. Around $40\%$ of the tidal disruption events are retrograde with respect to the initial disk angular momentum axis.
    \item The predicted mass accretion rates have a steep dependence on the eccentric debris disk mass, $\dot{M} \propto (\Mdisk)^2$. A debris disk of mass above $\SI{0.03}{\Mearth}$ is required to explain typical instantaneous mass accretion rates of $\SI{e7}{\g\per\s}$.
\end{enumerate}

The estimated fraction of white dwarf systems that exhibit eccentric debris disks, retrograde planetesimal populations, and tidal disruption events is dependent on the density profile of the pre-kick debris disk. Our results from Figure \ref{fig:tde_comp} show that our mechanism requires at least $\SI{0.03}{\Mearth}$ in the debris disk within the range $\SIrange{240}{480}{\AU}$ assuming a kick of $\SI{1}{\km\per\s}$. Throughout the letter, we have assumed that the natal kick is impulsive. In reality, the investigated range of semi-major axes lies in the complex transition region between planetesimals with $a < \SI{10}{\AU}$ where mass loss seems adiabatic, and planetesimals with $a > \SI{1000}{\AU}$ where the mass loss seems impulsive \citep{Stone2015}. Furthermore, stellar models predict that mass loss is pulsed \citep[e.g.,][]{Reimers1975, Bloecker1995, Bonsor2011} which adds more layers of complication. In the future, our work will consider the treatment of mass loss and the complex orbital evolution of planetesimals in response to a more realistic natal kick.

Additionally, our current work only considers a pre-kick circular disk with equal-mass particles. Moving forward, we plan to repeat this numerical experiment with a scattered disk-like distribution of planetesimals accompanied by one or more giant planets. Preliminary results have shown that the initial burst of tidal disruption events is more significant for the scattered disk but apsidal alignment is weaker. A follow-up study with long-term simulations is necessary to determine whether the eccentric debris disk and high tidal disruption rates are maintained in this case. We note that higher $N$ simulations are necessary to thoroughly explore debris disks with masses around $\SI{1}{\Mearth}$ or below. In future studies, we aim to extend our mechanism to polluted white dwarfs with cooling ages $\unit{\Gyr}$ or older, and make our claims about the tidal disruption event bursts more robust through longer and higher $N$ simulations.

\section{Acknowledgements} \label{sec:acknowledgements}

We gratefully acknowledge support from the Aspen Center for Physics and the organizers of ``Exoplanet Systems and Stellar Life Cycles: Late-Stage and Post-MS Systems". We are especially grateful to Melinda Soares-Furtado. We thank Meredith MacGregor and the anonymous referee for their valuable feedback which greatly improved the quality of this letter. SM gratefully acknowledges support from the Uplift Research Program at the University of Colorado Boulder. This work utilized resources from the University of Colorado Boulder Research Computing Group, which is supported by the National Science Foundation (awards ACI-1532235 and ACI-1532236), the University of Colorado Boulder, and Colorado State University. \\

\bibliographystyle{aasjournal}
\bibliography{ms}

\end{document}